\documentclass[utf8]{FrontiersinHarvard} 

\usepackage[onehalfspacing]{setspace}

\usepackage{url,hyperref,lineno,microtype}
\usepackage[onehalfspacing]{setspace}

\usepackage{amsmath,amsfonts}
\usepackage{algorithmic}
\usepackage{algorithm}
\usepackage{array}
\usepackage{textcomp}
\usepackage{stfloats}
\usepackage{verbatim}
\usepackage{graphicx}
\usepackage{cite}
\usepackage{xcolor} 
\usepackage{longtable}
\usepackage{lscape}
\usepackage{rotating}
\usepackage{colortbl}
\definecolor{Gray}{gray}{0.9}
\usepackage{array,multirow}
\newcolumntype{L}[1]{>{\raggedright\let\newline\\\arraybackslash\hspace{0pt}}m{#1}}
\newcolumntype{R}[1]{>{\raggedleft\let\newline\\\arraybackslash\hspace{0pt}}m{#1}}

\newcolumntype{C}[1]{>{\centering\arraybackslash}p{#1}}

\usepackage{hyperref}
\hypersetup{colorlinks=true}    
\hypersetup{colorlinks,
	citecolor=blue, 
	linkcolor=blue
}

\def\keyFont{\fontsize{8}{11}\helveticabold }
\def\firstAuthorLast{Lora-Millan {et~al.}} 
\def\Authors{Julio S. Lora-Millan\,$^{1,\dagger}$, Mahdi Nabipour,$^{2,\dagger}$, Edwin van Asseldonk,$^{2}$ and Cristina Bayón\,$^{2,\dagger,*}$}
\def\Address{$^{1}$Electronic Technology Department, Universidad Rey Juan Carlos, Madrid, Spain. \\
$^{2}$Department of Biomechanical Engineering, University of Twente, Enschede, The Netherlands\\
$^\dagger$These authors contributed equally to this work.}
\def\corrAuthor{Cristina Bayón, University of Twente, Faculty of Engineering Technology, Horst Complex  W119 P.O. Box 217, 7500 AE, Enschede, The Netherlands}

\def\corrEmail{c.bayoncalderon@utwente.nl}

\begin{document}

\onecolumn
\firstpage{1}

\title {Advances on mechanical designs for assistive ankle-foot orthoses} 

\author[\firstAuthorLast ]{\Authors} 
\address{} 
\correspondance{} 

\extraAuth{}

\maketitle







\begin{abstract}
Assistive ankle-foot orthoses (AAFOs) are powerful solutions to assist or rehabilitate gait on humans. Existing AAFO technologies include passive, quasi-passive, and active principles to provide assistance to the users, and their mechanical configuration and control depend on the eventual support they aim for within the gait pattern. In this research we analyze the state-of-the-art of AAFO and classify the different approaches into clusters, describing their basis and working principles. Additionally, we reviewed the purpose and experimental validation of the devices, providing the reader with a better view of the technology readiness level. Finally, the reviewed designs, limitations, and future steps in the field are summarized and discussed.

\tiny
 \keyFont{ \section{Keywords:} Ankle-foot orthosis, gait, passive, active} 
\end{abstract}

\section{Introduction}
Locomotion is a primary task for human beings and an essential component for a rich quality of life. There might be diverse (neurological or muscular) causes that limit the locomotion ability in humans, especially the efficiency and effectiveness of gait. Among all multi-body segments and muscles involved in walking, those related to the ankle joint are major contributors to perform the required mechanical work  \citep{Moltedo2018,Conner2022,Vaughan1999}.

Over the last decades, wearable assistive ankle-foot orthoses (AAFOs) have been developed and applied to assist ankle motion in humans. The main aim of these devices is to either reinforce and enhance the mobility in able-bodied subjects \citep{Moltedo2018}, or to restore, assist or rehabilitate lost functions of people with motor disorders \citep{Moltedo2018,Alam2014a, Bayon2023,Shorter2013}.

Despite the end goal to be achieved with the AAFO, a major distinction between these devices can be made according to their working principle. Passive AAFOs are those devices that rely on passive elements such as dampers or springs to store and release energy during gait, containing no control or electronics. Quasi-passive (or semi-active) AAFOs use computer control to adjust the performance of a passive element, and sometimes also hold a small motor to modulate their stiffness. Finally, active AAFOs, also called robotic exoskeletons, come with electric or pneumatic actuators connected to a power source to deliver assistive torques or forces to the users.
%

There are previous reviews that analyzed to some extent the application of articulated AAFOs and their influence on gait. These studies either focused on a single working principle such as passive \citep{Choo2021} or active \citep{Moltedo2018, Shorter2013}, or concentrated on a single gait impairment \citep{Alam2014a}. However, a more extensive comparison of different working principles (passive, quasi-passive, and active), and the effects when being used on both healthy and impaired subjects with different walking limitations is missing. 

The aim of this paper is to assess the mechanical designs of devices that focus on providing assistance at the ankle level, and evaluate the gait effects generated by these devices when being tested. This research effort is intended to:
\begin{itemize}
    \item Understand and describe the actuation mechanisms adopted within the three groups of AAFOs (passive, quasi-passive and active)
    \item Classify different robotic AAFOs with regard to their mechanical configuration, purpose and effects when evaluated in healthy and impaired subjects
    \item Summarize the novelties and limitations of the reviewed systems to be considered in future designs
\end{itemize}

In order to contextualize this work, we start with discussing the biomechanics of gait, highlighting the importance of the ankle-complex joint and giving an overview of the most common pathological gait patterns associated with ankle muscles' weakness and the demanded assistances. In section \ref{section:mechanical}, we classify different reviewed designs for AAFOs into clusters, reviewing their mechanical configuration and control. Subsequently, in sections \ref{section:assistance} and \ref{section:effects} we expose the extracted results on how the reviewed devices are applied in terms of the type of assistance provided and the effects generated on final users. Finally, we discuss the outcomes by providing input on advantages and disadvantages of the exposed mechanisms as well as proposing trends for future research.


\section{Human gait}
\begin{figure}[ht]
    \centering
    \includegraphics[width=98mm]{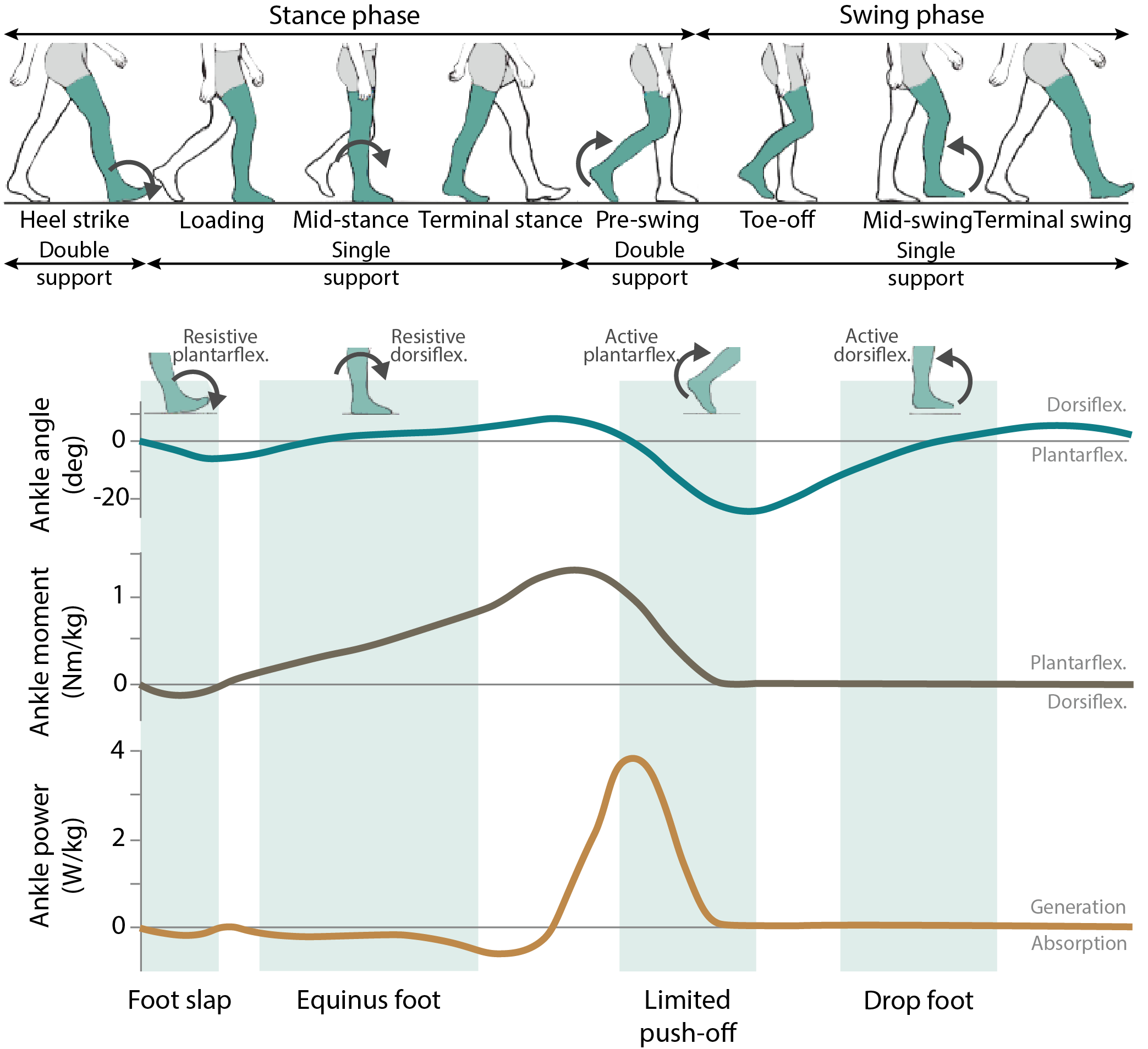}
    \caption{General overview of the human gait pattern. The ankle angle, ankle moment and ankle power normal patterns are shown along the gait cycle. In these graphs, the areas where the most common pathological gait patterns associated to the ankle-complex joint can be encountered are marked in green}. For each of these common impairments, the required ankle assistance is shown.
    \label{fig:Gait}
\end{figure}

Human gait is a very complex activity that involves the coordination of various systems such as the musculoskeletal, nervous, and cardiovascular systems \citep{Vaughan1999}. One of the main characteristics of healthy human gait is the high repeatability, and as such, it is normally segmented in cyclic gait patterns. A gait cycle starts with an initial heel contact with the ground and finishes with the following heel contact of the same limb (Figure~\ref{fig:Gait}). Two main phases take place within a gait cycle: the stance phase, in which the ipsilateral limb is in contact with the ground; and the swing phase, in which the ipsilateral foot raises from the ground and the limb advances for the preparation of the following heel strike \citep{Vaughan1999}.


\subsection{Importance of the ankle-complex joint}
The ankle-complex joint has an important and clear role during the full gait cycle \citep{Conner2022, Shorter2013}. At the beginning of the stance phase (initial heel contact), the muscles and tendons around the ankle are responsible for decelerating both the foot rotation before it enters in full contact with the ground, and the forward progression of the body. During mid-stance, the ankle-complex joint helps in maintaining balance and stability as all the user's weight is solely supported by the ipsilateral limb (while the contralateral limb is in the swing phase). Finally, with the forward tibia progression during stance and the contraction of ankle plantarflexor muscles, energy is stored to be later delivered at the moment of toe-off as plantarflexor torque. This accounts for a large exerted force that is the main source of forward propulsion power during gait \citep{Shorter2013} (Figure~\ref{fig:Gait}).

\subsection{Pathological gait patterns}
The role of the ankle-complex joint is compromised in many people with neurological disorders such as stroke, incomplete spinal cord injury or cerebral palsy. This causes a diminished walking ability and the appearance of pathological gait patterns \citep{Armand2016, Beyaert2015, Wirz2018}. The impairments at the ankle level can be related to reduced strength, muscle dysfunction, spasticity, poor selective control or limited range of motion (ROM). Due to the importance of the ankle during gait and in order to limit the consequences of its impaired function, AAFOs are the foremost used type of assistive devices to support walking limitations \citep{Conner2022, Bayon2023}. For a proper design of AAFO devices, it is key to get a good understanding of the effects of the different pathological gait in the subjects' walking patterns \citep{Shorter2013, Conner2022} (Figure~\ref{fig:Gait}).

Weakness in the ankle dorsiflexors (primarily represented by the tibialis anterior) may affect gait in both swing and initial stance phases \citep{Shorter2013, Rodda2001}: during swing, the most common gait deviation is the \textbf{drop-foot}, reflected by an insufficient toe clearance. This deficit is normally compensated by increased knee and hip flexion and hip abduction during swing. Additionally, during initial stance and just after initial heel contact, dorsiflexors' weakness can also prevent the controlled deceleration of the foot, which is often known as \textbf{foot-slap}.

An impaired function of ankle plantarflexors (with the gastrocnemius and soleus muscles as main contributors) mostly affect the stance phase of gait. Especially, the existence of spasticity or contracture in the gastrocnemius and/or soleus results in \textbf{equinus pattern} \citep{Rodda2001,Brockett2016}, characterized by an abnormal (increased) ankle plantarflexion through most of the stance phase. Additionally, ankle plantarflexors dysfunction also causes a \textbf{limited push-off} power due to the inability of the affected muscles to concentrically contract at the end of the stance phase \citep{Brockett2016}. In general, weakness of ankle platarflexors affects stability, reduces walking speed, shortens step length, and increases the energy cost of walking.

\subsection{Type of ankle assistances}
Although the ankle-complex joint has three degrees of freedom (DOF), the primary plane during locomotion and where most of the motion is performed is the sagittal plane (Figure~\ref{fig:Gait}). Ideally, AAFOs designed to assist walking in neurological disorders should be able to accommodate dorsi- and plantarflexion assistances depending on the user's needs. This is translated into different assistive strategies related to the targeted gait impairment: (1) avoiding excessive ankle plantarflexion during swing that causes drop-foot, (2) controlling deceleration of the foot at initial heel contact to inhibit foot-slap, (3) preserving stability during mid-stance while preventing equinus pattern, or (4) generating an assistive plantarflexion torque at the instant of push-off. The last one is also a common strategy used to enhance gait in able-bodied subjects.

The accomplishment of all the above-mentioned actions within a design is a real challenge, especially because adding extra weight to a device located at the ankle (distal joint) could hamper the performance of the user increasing their energy cost of walking \citep{Bayon2023}. Moreover, if the purpose of the AAFO device is to be used during daily-life activities, the design should allow some adaptability to the challenging mobility tasks and ground variations encountered in everyday use (e.g. stairs, ramps, uneven terrains...) \citep{Bayon2023}.

\section{Mechanical designs for AAFOs}
\label{section:mechanical}
\subsection{Passive AAFOs}

Passive AAFOs are those that do not require any external power supply to generate the assistance. They have the main benefit of being lightweight compared with active AAFOs, so their use is closer to daily basis. To provide assistance, passive AAFOs need to be articulated, and normally they rely on passive elements such as springs, dampers, or elastic bands (conventional non-assistive passive AFOs are out of the scope of this manuscript). In most cases, the design parameters of the passive elements (e.g., the stiffness of the spring) can be optimized only in very specific situations. Thus, these devices have the main limitation of hindering users' performance for tasks they are not designed for. Regardless of the passive element that is being used, a relevant classification of passive AAFOs can be done based on their working principle and the mechanical configuration of the device (see Supplementary Materials).

\subsubsection{Continuous adjustable dynamic response}
\begin{figure}[ht]
    \centering
    \includegraphics[width=170mm]{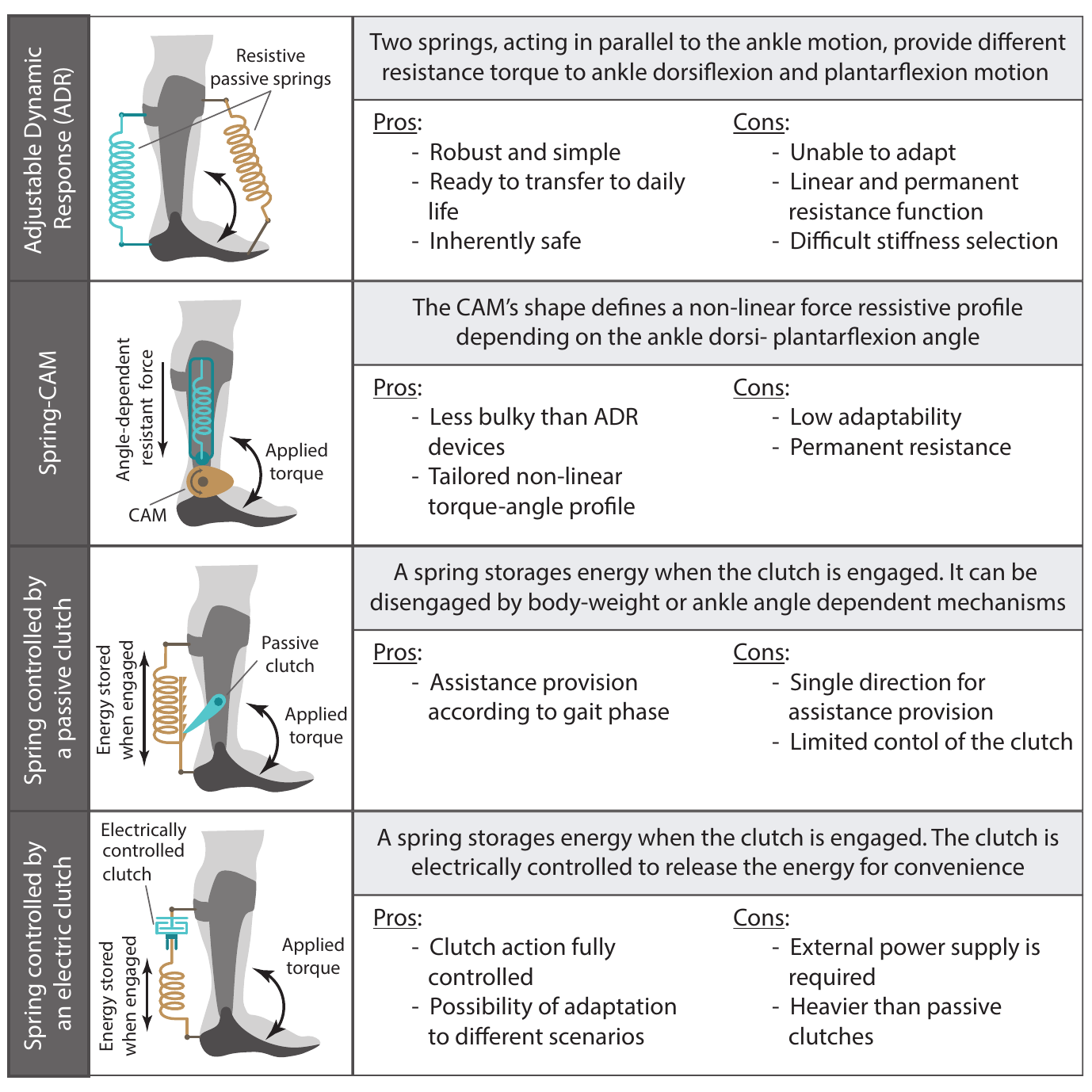}
    \caption{Clusters classification for the main mechanical configuration of passive and quasi-passive AAFOs. The working principle and main pros and cons are given for: (a) continuous adjustable dynamic response AAFO; (b) Spring-cam AAFO; (c) clutch-dependent AAFO; (d) Electrically-controlled clutch dependent AAFO.}
    \label{fig:Passive}
\end{figure}

The importance of tuning the mechanical properties of an articulated AAFO has gained importance in the last few years, and in that regard, adjustable dynamic response AAFOs (ADR-AAFOs) (Figure~\ref{fig:Passive}) have been introduced in the market as a potential solution \citep{BeckerOrthopedic2018,Ottobock2018,FiorandGentz2021,UltraflexSystems2009}. ADR-AAFOs allow for a variable ankle ROM and selective dorsi- and plantarflexion resistance during both stance and swing phases, making these devices suitable for assisting the full gait cycle. Normally, the working principle of the ADR-AAFOs is based on the use of two springs, one for dorsiflexion and one for plantarflexion. However, this can be enhanced by adding additional springs in parallel that activate at different instants \citep{kobayashi2017}, simulating the effects of a variable stiffness actuator \citep{Vanderborght2013}. The torques of different designs can be adjusted either by replacing the assistive springs or by modifying the spring preload. Nevertheless, once the spring stiffness has been selected, the provoked torque affects the user along the whole gait cycle, eliminating the possibility for online modulations.

The target users' group of ADR-AAFOs is very wide \citep{Kerkum2015,Meyns2020,Wren2015,Kobayashi2019}, but all the designs have the main common intention of dynamically bringing the users into an upright position and improving their stability while walking and standing. This gain on stability may be related to the support of one or more ankle impairments encountered within the gait cycle (Figure~\ref{fig:Gait}).


\subsubsection{Spring-cam}
Spring-cam mechanisms can be used to compensate for diverse walking impairments by customizing individualized torque-angle curves according to the cam shape and the spring stiffness (Figure~\ref{fig:Passive}). The cam profile changes the moment arm, which effectively results in a characteristic torque-deflection mechanism that could be also considered to have variable stiffness around the ankle \citep{Vanderborght2013}. Thus, they provide personalized dorsi- and plantarflexion torques to fit the user's needs. An example of a spring-cam prototype is presented in \citep{Sekiguchi2020}, where authors designed a solution to store energy during dorsiflexion (mid and late stance) that is later released to compensate for push-off (plantarflexion). This mechanism was incorporated in the AAFO device developed in \citep{Yamamoto2005}. 

In patients suffering from equinus pattern, the self generated activity of the human has a large effect on the ankle motion. Rodríguez et al. \citep{Rodriguez2018} proposed a spring-cam AAFO to compensate for the increased human stiffness in these patients. The proposed solution used a negative stiffness mechanism composed by a pre-loaded gas-spring and a cam follower.

\subsubsection{Clutch-dependent}
Clutch-dependent AAFOs (Figure~\ref{fig:Passive}) offer the possibility of selecting the periods within the gait cycle in which the stiffness of the passive element should be present. These clutches can be passively controlled, and a common way of doing so is based on either the user's body weight \citep{VanHoorn2022,Yandell2019,Wang2020a,Liu2021,Hirai2006} or the ankle angle during gait \citep{Leclair2018,Collins2015,Pardoel2019}.

Body-weight dependent clutches for AAFOs are normally located at the bottom of the user's shoe, and they use the weight of the user during the stance phase to passively control the storage and release of energy. The core of a body-weight dependent clutch can be either compliant (e.g., pneumatic) \citep{Hirai2006}, a mechanism \citep{Wang2020a, Liu2021}, or consisting of a slider-spring system \citep{Yandell2019, VanHoorn2022}. Several of these approaches have been designed for supporting push-off \citep{VanHoorn2022,Yandell2019,Wang2020a,Liu2021} by engaging an elastic spring when the ankle is dorsiflexing during the stance phase and releasing the stored energy when the ankle starts to plantarflex during push off, without hindering the ankle motion during swing. Other designs had the goal of preventing drop-foot by locking the ankle joint in a neutral position during the swing phase \citep{Hirai2006}. In this last case, the ankle motion is only constrained during swing and not during the stance phase \citep{Hirai2006}.

Ankle angle dependent clutches for AAFOs use the motion of the ankle joint during gait to control the storage and release of energy. These mechanisms are normally more complicated than body-weight clutches as, during a gait cycle, the ankle goes through plantarflexion twice, and thereby, this motion is not directly related to stance and swing phase (as happened with the body-weight). In this case, the pawl-ratchet principle is often used to engage and disengage the mechanism. Angle dependent clutches can be connected to passive elements such as springs \citep{Collins2015}, or to pneumatic artificial muscles (PAMs) that do not require an external power source \citep{Leclair2018,Pardoel2019}. Devices based on this approach normally have the final goal of empowering human gait by assisting push-off. 

\subsubsection{Others}
Oil dampers can also be used as hydraulically resistant shock absorbers. When the ankle joint moves to plantarflexion during initial stance the resistance produced by an oil damper reduces the rate of angular change. Following this working principle, Yamamoto et al. \citep{Yamamoto2005} developed a passive AAFO to satisfy the requirements for patients with hemiplegia by improving the insufficient eccentric contraction of the dorsiflexors at initial stance (preventing foot-slap). 

\subsection{Quasi-Passive AAFOs}
Passive AAFOs are less adaptable than active (powered) devices due to their simplified design and their inability to control their functions. A trade-off between both solutions are the quasi-passive (or semi-active) devices, where sensors and small motors with low energy consumption are used to modulate the stiffness of passive elements (Supplementary Materials). This enhances the adaptability of the devices to changes in gait conditions or environments, allowing the passive element to work in a closed-loop to some extent. 

\subsubsection{Electrically-controlled clutch}
Similar to passive AAFOs, there are designs for quasi-passive devices that rely on the action of a clutch. However, in this case, the clutch is normally electrically controlled, and sometimes a small motor is also used to combine both passive energy storage and additional power supply (Figure~\ref{fig:Passive}). Such is the case for the design of Zhang et al. \citep{Zhang2015}, where an electromagnetic clutch was used to command the storage of energy in a compression spring. A DC motor added additional power for energy storage, which offered, in a sense, the inherent variable stiffness \citep{Vanderborght2013}. This device was intended to absorb heel strike impact and enhance push-off.

Another electromagnetically driven clutch-dependent AAFO designed to capture heel-strike energy loss and recycle it into push-off propulsion was recently presented by Wang et al. \citep{Wang2022}. The clutch and the elastic energy storage mechanism are coupled together by a stiff assistance spring. 

Finally, Defman et. al. \citep{Dezman2018} presented the modification of an originally passive AAFO making it quasi-passive through an electrically controlled ratchet-pawl clutch. By including the active clutch, the authors could solve the problems with timing of engagement and disengagement specific to each user, being more reliable than with the original passive clutch. 

\subsubsection{Others}
Other quasi-passive designs have been proposed in the literature with the same aim of improving the possibilities of passive AAFOs by adjusting the stiffness or damping. As these designs vary quite a lot from each other, we provide a short description of four of them without following a specific classification.

The first solution, called SmartAFO \citep{Oba2019a}, used a quasi-passive technology for the mitigation of drop-foot and foot-slap during gait. In particular, the AAFO used a magnetorheological fluid damper to selectively modulate a variable impedance around the ankle joint. Equipped with an elastic link, the system allows energy storage and release according to its coil current.

Kumar et al. developed a quasi-passive AAFO for stiffness auto-tuning and adaptability across different walking conditions \citep{Kumar2020}. In this solution, a stepper motor is used to vary the stiffness of a passive spring in real-time, while electromyography data of the user is being recorded for closing the control loop.

Diller et al. designed a lightweight, low-power electroadhesive clutch to demonstrate its usage for adjustable spring stiffness in AAFOs \citep{Diller2016a}. The authors placed several clutched springs in parallel to discretely produce six different levels of stiffness. However, the main drawback of electroadhesive clutches is their failure because of dielectric breakdown, which can expend a significant amount of electrical energy and cause the clutch to disengage.

Finally, the quasi-passive AAFO designed in \citep{Allen2021} is a dielectric elastomer actuator (DEA) that connects the foot near the toes to the shank under the knee. DEAs are a kind of artificial muscle that, when charged with a constant voltage, produce electrostatic forces. The DEA passively keeps the foot dorsiflexed to the maximum angle recorded during normal gait, making it suitable for drop-foot patients. This device automatically adapts to the user's gait due to the gait phase detection ability of this AAFO.

\subsection{Active AAFOs}
There are certain conditions of motion control and assistance that cannot be accomplished with passive elements due to their limited adaptability and control. Active AAFOs have the advantage of using external power supplies and actuators to address these limitations. Differences between active AAFOs' designs can be found looking to their mechanical configuration and the control strategies used to assist ankle function (see Supplementary Materials).

\subsubsection{Mechanical configuration of active orthoses}

\begin{figure}[ht]
    \centering
    \includegraphics[width=170mm]{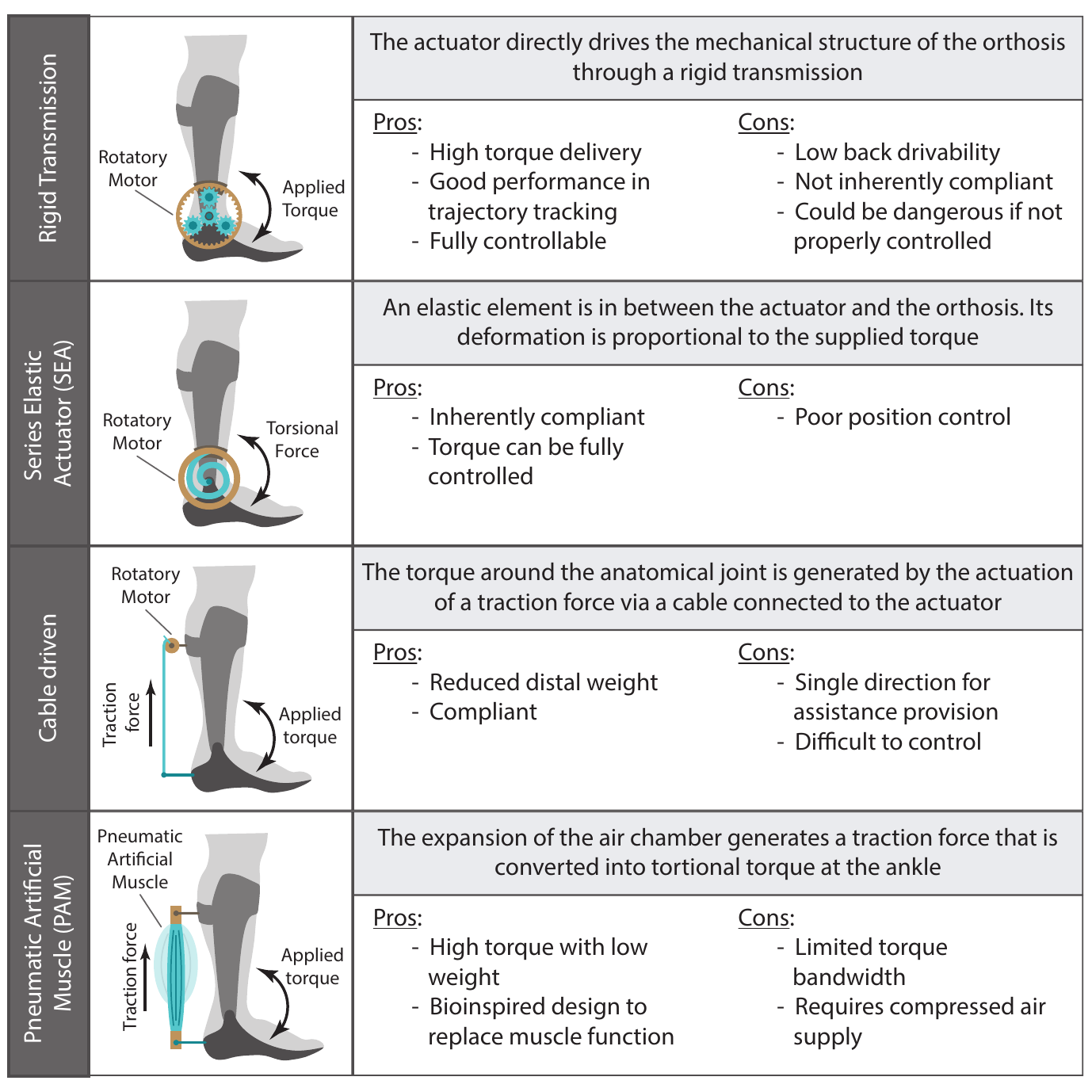}
    \caption{Clusters classification for the main mechanical configuration of active AAFOs. The working principle and main pros and cons are given for: (a) AAFO with rigid transmission; (b) AAFO with elastic element -SEA- in the transmission; (c) AAFO driven by a cable; (d) AAFO with a pneumatic artificial muscle -PAM-.}
    \label{fig:Active}
\end{figure}

    \paragraph{Rigid transmission}
    The straightforward configuration is composed of a rigid transmission between the actuator and the orthosis structure \citep{Choi2018, DeMiguel-Fernandez2022a, DeMiguel-Fernandez2022b, Moosavian2021, ISI:000510675700010, Kim2020, Li2011, Shorter2011, Yeung2018, Yeung2017, Yeung2021} (figure~\ref{fig:Active}). This configuration aims at ensuring the proper delivery of the torques and forces generated by the actuators. However, as a disadvantage, it is not inherently compliant, as it does not allow adaptations to avoid possible misalignment. Moreover, the movement of the mechanism is restricted to the one commanded by the controller.
    
    \paragraph{Series elastic actuator}
    To improve the adaptability, comfort, and efficiency of active AAFOs, some research groups have opted for using compliant elements in the torque transmission. These elements allow certain adaptations to the user and reduce the movement restrictions imposed by the orthoses. In this regard, a common approach is the use of elastic elements in series with the actuators to achieve the mentioned compliance \citep{Roy2013, ISI:000377575200001, Kim2011, ISI:000220173200003, ISI:000295363000003, Oymagil2007, VanDijk2017, Mooney2016, Tamburella2020, Meijneke2014a, Dzeladini2016, Bayon2022, Durandau2022} (see figure~\ref{fig:Active}). This configuration, also called Series Elastic Actuator (SEA), not only enables a less restrictive mechanism, but also allows the direct control of the exerted torque or force by controlling the deformation of the elastic element. Moreover, the SEA configuration allows for the implementation of a variable stiffness control to better adapt to natural changes of muscle tension \citep{Vanderborght2013}.
    
    \paragraph{Cable driven}
    A different approach to obtain a compliant mechanism while reducing the inertia of the device is the use of cables or wires to transmit the force to the human segments (figure~\ref{fig:Active}) \citep{Mooney2014,Mooney2014a,Witte2015,Bae2015,Jackson2015,Awad2017b,Awad2017a,Steele2017,Zhang2017c,Bae2018,ISI:000446394502033,Lerner2018,McCain2019,Bougrinat2019,Jackson2019,Gasparri2019,Lerner2019a,Siviy2020,Orekhov2020,Xia2020,Gomez-Vargas2021,Conner2021a,Harvey2021,Orekhov2021,Acosta-Sojo2022,Wang2022h,Peng2022b,Fang2022a,Fang2022b,Chen2022, Slade2022}. These devices allow the placement of the major part of their weight in a proximal location (e.g. the pelvis), reducing the mass placed in the distal segments and therefore the inertia added by the device. In addition, some of them do not require a fixed rotatory axis at the AAFO (e.g. \citep{Bae2015, Mooney2016}), negating the need of a precise alignment between the human ankle and the AAFO joint. 
    
    \paragraph{Pneumatic artificial muscles}
    Following a bioinspired approach, PAMs were developed to generate forces when they get deformed by pressurized air. Some researchers opted for the use of PAM actuators for active AAFOs \citep{Gordon2007,Cain2007,Sawicki2008,Kinnaird2009,Kao2010b,Galle2013,Galle2014,Koller2015,Takahashi2015,Galle2015,Koller2017,Galle2017,Malcolm2018} (Figure~\ref{fig:Active}). These actuators enable a high torque supply with a lightweight configuration. However, they are also characterized by slow dynamics while providing assistance and they require a pneumatic pump to power them.


\subsubsection{Control of active orthoses}
The performance of active AAFOs is determined by the control strategy followed to supply the intended assistance. In this regard, the control algorithm of these devices can be analysed following the bio-inspired hierarchical paradigm previously reported \citep{Tucker2015, Baud2021}. This paradigm divides the control of robotic exoskeletons into three levels, similarly to the control of human movement by the nervous system. The high-level control corresponds to the volitional control of the movement performed by supraspinal structures. The mid-level control is responsible for the generation of cyclic movement patterns by the spinal Central Pattern Generators. In the lowest level, the afferent information is integrated by the nervous system to properly generate the efferent signal to activate and deactivate agonist and antagonist muscles to effectively perform the movement.

In the context of wearable robotic devices, particularly active AAFOs, these neural controllers are translated into control algorithms. The high-level controllers are responsible for the intention detection and the context extraction, being independent of the device itself. For this reason, we do not report the high-level control in this manuscript. The mid-level controllers, responsible for the movement pattern generation, and the low-level controllers, responsible for closing the feedback loop, are described below.

\paragraph{Low-level control}
Depending on the low-level control of active AAFOs, the deliverance of the assistance and thus the actuator behaviour can be performed following different paradigms. The most used approach is based on torque or force closed-loops \citep{Choi2018, Kim2020, Peng2022b, Yeung2021, Lerner2018, Orekhov2020, VanDijk2017}. These controllers have the advantage of being directly involved in the delivery of the robotic assistance by counteracting the muscular weakness of the impaired users. Other approaches use impedance \citep{ISI:000377575200001} or position \citep{ISI:000510675700010} controllers to command the assistance (see Table S4 in the Supplementary Materials). 
    
\paragraph{Mid-level control}    
Mid-level controllers are also required to coordinate an AAFO's action with the current user's gait state, thus ensuring that the assistance is delivered at the correct timing (Table S4 in the Supplementary Materials). This coordination is crucial to avoid instabilities and maximize the benefits of wearing the active device. Mid-level controllers have been previously assessed in a general way for partial and unilateral robotic exoskeletons \citep{Lora-Millan2022}, but without focusing on ankle devices.
The most common control strategy to achieve a proper coordination is based on the detection of several gait events that divide the gait into different phases. For each gait phase, an impedance model can be applied to set the assistance level \citep{ISI:000377575200001, Kim2020}. Additionally, the detection of gait events can also be used to trigger the execution of assistive torque patterns \citep{Mooney2014a, Acosta-Sojo2022}. 

Since these strategies are limited to the instants of detection of the key events, some authors prefer to characterize the full step with a monotonically increasing signal, called continuous gait phase, that is reset once the step is over. Thereby, it enables a more versatile control strategy not limited to key instants. The gait phase can be calculated based on the duration of previous steps and the time elapsed from the last heel-strike \citep{Bougrinat2019} or can be estimated in real-time by using adaptive frequency oscillators \citep{VanDijk2017}. 

Finally, a third option would be to generate reference trajectories to be followed by the device from data directly measured on the subject. These references can be based on movement information from the assisted joint \citep{Lerner2019a, Harvey2021} or on the muscular activity that generates such movements \citep{Koller2015, McCain2019}.
    

\section{Type of assistance provided}
\label{section:assistance}
Although the reviewed literature (tables in Supplementary Materials) has revealed different mechanical designs for passive, quasi-passive, and active devices, a possible combined classification of all of them can be done according to their goal: (1) to empower healthy walking, (2) to assist pathological gait, or (3) for rehabilitation purposes. Additionally, within each general purpose, we can identify the type of walking assistance that every device is intended for. These results for all analyzed devices (a total of 54) are presented in Table \ref{table:purposeDistribution}. Although the majority of these devices were designed for a unique target population (healthy or impaired walking subjects), some of them considered the gait assistance for both populations \citep{Leclair2018,Yandell2019, Wang2022, Bayon2022}.

\renewcommand{\arraystretch}{1.5}
\begin{table}[t]
\centering
\caption{Distribution of type of assistance and purpose (empowerment/assistive or rehabilitation) across passive, quasi-passive and active devices. Note that some devices are intended to support both impaired and healthy users. The individual publications making up these total numbers can be checked in the Table S1 of the Supplementary Materials.}
\label{table:purposeDistribution}
\resizebox{\columnwidth}{!}{%
\begin{tabular}{ccccccccc}
\multicolumn{1}{l}{} &
  \multicolumn{1}{l}{} &
  \multicolumn{1}{l}{} &
  \multicolumn{6}{c}{Type of assistance} \\ \cline{4-9} 
 &
   &
   &
  Drop-foot &
  Foot-slap &
  Push-off &
  Full gait &
  Drop-foot + foot-slap &
  Drop-foot + push-off \\ \hline
\multicolumn{1}{c|}{\multirow{5}{*}{Passive}} & Impaired assist. & \textbf{11} & 3 & 1 & 3  & 4  &   &   \\
\multicolumn{1}{c|}{}                         & Healthy assist.  & \textbf{3}  &   &   & 3  &    &   &   \\
\multicolumn{1}{c|}{}                         & Both assist.     & \textbf{2}  &   &   & 2  &    &   &   \\
\multicolumn{1}{c|}{}                         & Rehabilitation    & \textbf{}   &   &   &    &    &   &   \\ \cline{2-9} 
\multicolumn{1}{c|}{}                         & TOTAL    & \textbf{16} & 3 & 1 & 8  & 4  &   &   \\ \hline
\multicolumn{1}{c|}{\multirow{5}{*}{Quasi-passive}} &
  Impaired &
  \textbf{2} &
  1 &
   &
   &
   &
  1 &
   \\
\multicolumn{1}{c|}{}                         & Healthy assist.  & \textbf{3}  &   &   & 3  &    &   &   \\
\multicolumn{1}{c|}{}                         & Both assist.     & \textbf{1}  &   &   & 1  &    &   &   \\
\multicolumn{1}{c|}{}                         & Rehabilitation    & \textbf{}   &   &   &    &    &   &   \\ \cline{2-9} 
\multicolumn{1}{c|}{}                         & TOTAL    & \textbf{6}  & 1 &   & 4  &    & 1 &   \\ \hline
\multicolumn{1}{c|}{\multirow{5}{*}{Active}}  & Impaired assist. & \textbf{19} & 2 &   & 5  & 7  & 1 & 4 \\
\multicolumn{1}{c|}{}                         & Healthy assist.  & \textbf{10} &   &   & 10 &    &   &   \\
\multicolumn{1}{c|}{}                         & Both assist.    & \textbf{1}  &   &   &    & 1  &   &   \\
\multicolumn{1}{c|}{}                         & Rehabilitation    & \textbf{2}  &   &   &    & 2  &   &   \\ \cline{2-9} 
\multicolumn{1}{c|}{}                         & TOTAL    & \textbf{32} & 2 &   & 15 & 10 & 1 & 4 \\ \hline
\end{tabular}%
}
\end{table}


Assisting paretic ankle functions is the most common objective of AAFOs, being the purpose of 36 out of 54 reviewed devices (considering passive, quasi-passive, and active AAFOs). The main difference between devices' approaches relies on the strategy adopted to assist the impaired walking. Some devices (12 out of 36) were designed to provide assistance during the whole gait cycle, e.g. \citep{Meyns2020, Bae2018}. The rest of them were aimed to assist concrete gait functions, thus focusing on push-off (11 out of 36, e.g. \citep{Leclair2018, Takahashi2015}), drop-foot (6 out of 36, e.g. \citep{Rodriguez2018, Kim2020}) or foot-slap (1 of of 36, e.g. \citep{Yamamoto2022, Kim2011}). Other authors assisted more than one gait function at the same time, for instance, drop-foot and foot-slap (2 out of 36, e.g. \citep{Kim2011}) or drop-foot and push-off (4 out of 36, e.g. \citep{Xia2020}).

When authors pursued to improve healthy performance (20 out of 54), they always opted for complementing the push-off function of the ankle, regardless whether they followed passive, e.g. \citep{Wang2020a, Yandell2019}, quasi-passive, e.g. \citep{Wang2022} or active approaches,  e.g. \citep{Koller2017, Mooney2016}.

Finally, some AAFOs were designed to be used for rehabilitation practices. This objective represents the least common purpose, being only the function of 2 out of 54 devices \citep{Yeung2021, Li2011}. These two devices are active AAFOs designed under the rigid transmission paradigm. They are able to act over the whole gait cycle, actuating in both dorsi- and plantarflexion movements.

\section{Human-robot-environment interaction}
\label{section:effects}
\subsection{Adaptability of the devices}
Adapting to different gait phases, speeds, and terrains is of a great significance for any AAFO. Limiting these adaptations would imply a reduced versatility of the device as well as a hindrance towards the user's improvement. Moreover, adaptability of the AAFOs does not only refer to adaptations to the environment or walking condition, but also the required adaptation to the user's progression while using the device.

Passive AAFOs that do not utilize clutches have inherent limitations in terms of adaptability, as the open-loop function of passive elements limits their ability to adapt to different walking conditions and tasks. These passive devices are not capable of providing assistance to either dorsi- or plantarflexion without hindering the other. The problem is resolved to some extent by using body-weight dependent clutches (e.g. \citep{VanHoorn2022,Yandell2019}). These types of clutches can detect the switch between different phases of gait regardless of the speed of the user, but their effectiveness has not been tested yet when walking over different terrains.

Quasi-passive AAFOs can more accurately switch the function of the passive elements to better react to changes in gait phases \citep{Wang2022}. Some of them can adjust the stiffness of the AAFO joint to better match the stiffness of the biological ankle \citep{Allen2021}, or directly adjust the assistance provided by the device on the subject \citep{Oba2019a, Allen2021, Diller2016a}.

Finally, active AAFOs are able to control dorsi- and plantarflexion separately, thereby, they offer more possibilities for adaptation irrespective of whether their main purpose is to support drop-foot (e.g. \citep{Kim2020, ISI:000220173200003, Awad2017b}), foot-slap (e.g. \citep{Kim2011, Choi2018, Chen2022d}), push-off (e.g. \citep{Choi2018, Acosta-Sojo2022, ISI:000295363000003}) or full gait (e.g. \citep{Lerner2018, Siviy2020, DeMiguel-Fernandez2022a}). In fact, apart of being able to adapt to walking velocities, they also have the ability to control their actuator to be active during some phases of the gait and transparent during other phases \citep{Bayon2022}.

\subsection{Effects on users}
Regarding the subjects involved in the experiments, three main validation strategies have been reported across the reviewed articles (Figure \ref{fig:experimentalValidation}). Some authors just reported technical validations regarding the proper functioning of the device and the assistance generated in terms of force and torque (e.g. \citep{Leclair2018, Diller2016a, Choi2018}) but without directly testing in subjects. When subjects were involved in the experiment, we can distinguish between those that involved healthy (e.g. \citep{Wang2020a, Oba2019a, Malcolm2018}) or impaired subjects (e.g. \citep{Sekiguchi2020, Fang2022a}). Remarkably, the AAFOs considered in this manuscript are more extensively validated with healthy than with impaired walking subjects. This may be due to two main reasons: (1) some AAFOs are directly developed to empower healthy subjects and increase their endurance while walking; (2) healthy subjects are considered as a first step for the validation of AAFOs, even if they are designed to assist paretic patients. 


When reporting validations with subjects, authors focused on the assessment of three main aspects: (1) the lower-limbs kinematics, related to functional aspects of gait; (2) the muscular adaptations that occur when users received the assistance provided by the AAFO; and (3) the metabolic cost of walking.

\subsubsection{Effects on kinematics}
\begin{figure}[ht]
    \centering
    \includegraphics[width=88mm]{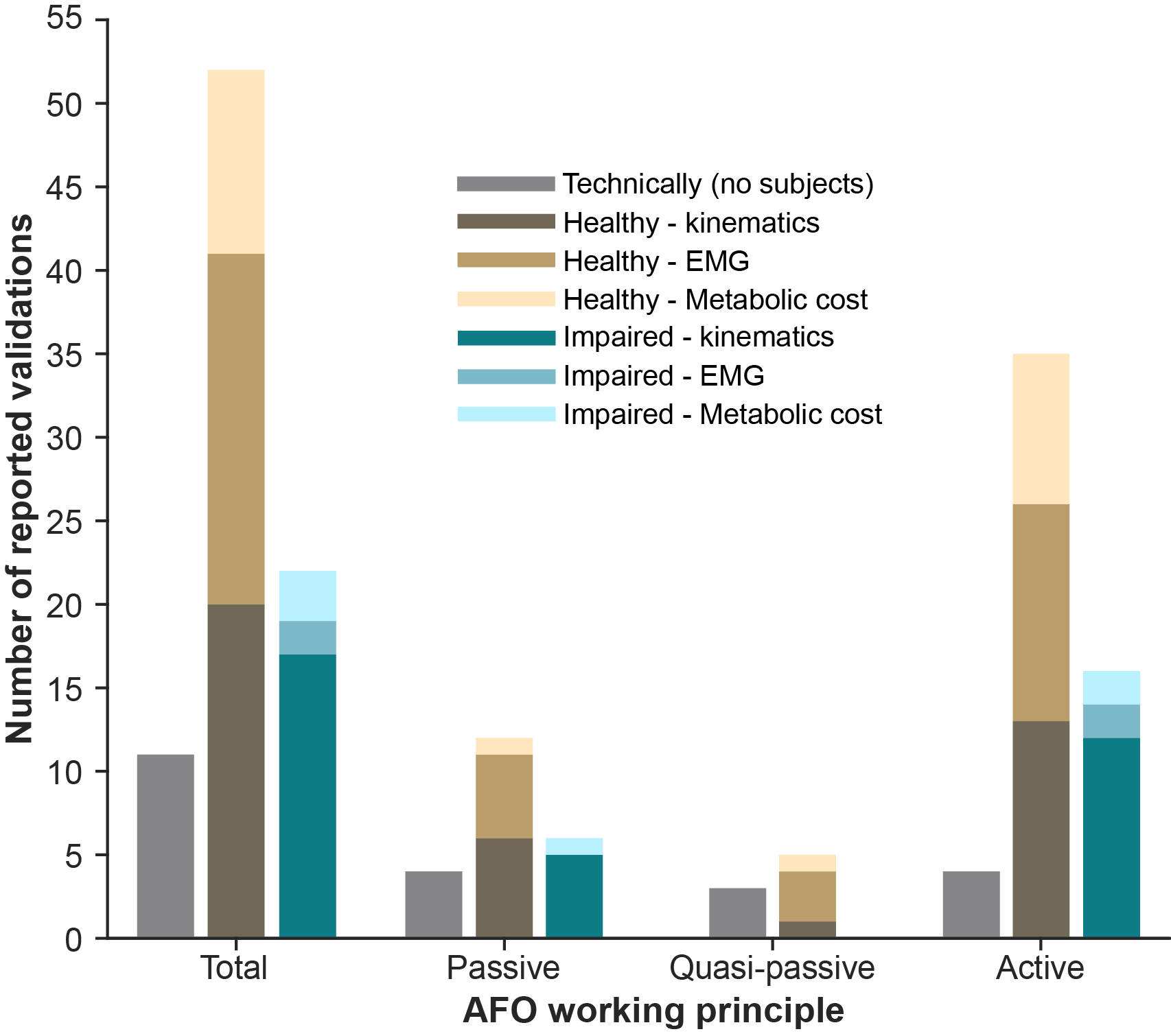}
    \caption{Distribution of experimental validations with AAFOs reported in the literature.}
    \label{fig:experimentalValidation}
\end{figure}

Kinematic analysis is the most common assessment when impaired walking subjects are involved (17 out of 22 validations with impaired subjects reported it, Figure \ref{fig:experimentalValidation}). Although quasi-passive AAFOs have not been validated in this sense, passive and active devices have shown their feasibility in assisting and improving impaired gait kinematics. Reported results included improvements of the ROM, e.g. \citep{Wren2015, Sekiguchi2020, ISI:000220173200003, Kim2020}, reduction of gait deficits such as drop-foot, e.g. \citep{Roy2013, Yeung2017}, foot-slap, e.g. \citep{Kim2011}, or inter-limb asymmetry e.g., \citep{Bae2015, Awad2017b}. Additionally, others also showed kinetics effects as an increase of the torque on the paretic side \citep{McCain2019}. 

Remarkably, kinematics are being less assessed in evaluations with healthy subjects (not impaired), only accounting for 20 out of 52 of the reported validations with humans (Figure \ref{fig:experimentalValidation}). Those that performed kinematic analysis in healthy subjects focused on increasing gait speed \citep{Slade2022} or stability \citep{Gonzalez2022b, Bayon2022}, for example.

\subsubsection{Effects on EMG}
Effects in muscular activity is the most common outcome reported when healthy subjects are involved in the experiment (21 out of 52 validations, Figure \ref{fig:experimentalValidation}). From these reported assessments, 5 out of 21 were with passive devices, 3 out of 21 with quasi-passive, and 13 out of 21 with active. Generally, the evaluations using EMG with healthy reported a reduction of activity of muscles involved in the ankle movement (e.g. soleus \citep{Chen2022, Yandell2019}, gastrocnemius \citep{Acosta-Sojo2022, Bougrinat2019} or tibialis anterior \citep{Rodriguez2018, Weerasingha2018}). 

When impaired subjects were involved, EMG effects were reported only for active devices (e.g. \citep{Orekhov2020, Fang2021}), with main reductions in soleus activity.

\subsubsection{Effects on energy expenditure}
Several AAFOs have been focused on increasing the endurance and reducing the mechanical power of healthy or impaired subjects while walking \citep{Mooney2016, Koller2017}. In this sense, the most common assessment is to evaluate the metabolic consumption of the user walking while being assisted by the tested AAFO. 

When healthy subjects were involved, the energy expenditure was assessed for all kinds of devices (e.g. passive \citep{Collins2015}, quasi-passive \citep{Wang2022} and active \citep{Shepherd2022}). Irrespective of the device used,  a reduction in the metabolic cost of walking was observed in most cases. However, when assisting impaired subjects, this evaluation was only done for active AAFOs \citep{Orekhov2020, Jackson2019}. Also, these studies reported a reduction in the energy consumption.

\section{Discussion}
%
In this manuscript, passive, quasi-passive, and active AAFOs have been identified and further categorized based on their working principle and mechanical configuration. As a result, three clusters were created to enclose passive devices, one for quasi-passive devices and four for active devices. Additionally, we discussed the possibilities of these devices to provide assistance, to adapt to the environment, and their effects when being tested with final users. 

There is an extreme difficulty in quantitatively comparing devices' performances and the effectiveness of the different solutions among each other. The primary reason is the varying and limited number of test subjects involved in the evaluations. The majority of designs have shown their effects only on a small number of healthy subjects, and evaluations on impaired subjects are even more scarce. Moreover, researchers do not have a defined consensus or a set of criteria for evaluating their devices following similar validation procedures, which makes it even more challenging to quantitatively compare the performances. Therefore, we focus here on discussing differences in designs and preferred devices' applications, identifying the advantages and disadvantages of each cluster individually, and proposing a set of criteria that would be useful to consider in the future to ensure consistent evaluations.

\subsection{Design-related advantages and disadvantages}
\label{sec:AdvDisadv}
AAFOs are always intended to assist ankle motion based on the specific user's needs, which is done by the provision of assistive ankle torques. The main reason to select an AAFO design over others is the ability of the device to tailor the assistance to address those specific needs.

The principal design-related benefit of passive and quasi-passive devices compared with active ones is their reduced weight, as they do not use heavy actuators or batteries. From these, ADR-AAFOs are the passive devices closer to the market. They support the ankle during the entire gait cycle, providing bidirectional torques by means of passive springs \citep{BeckerOrthopedic2018,Ottobock2018,FiorandGentz2021,UltraflexSystems2009}. However, these devices are reported to be still bulky, and it is difficult to adjust their spring stiffness to the specific user's pathological gait \citep{Kerkum2014, kobayashi2017}. Moreover, the fact of providing a constant stiffness along the whole gait cycle without a dedicated timing for the storage and release of energy, may hinder the subject's residual capabilities. This, together with the lack of evidence that proves their ability to adapt to different scenarios and mobility tasks encountered in everyday life, can be considered as the big limitations for these solutions \citep{Firouzeh2021, Kerkum2015,Bayon2023}.

Spring-cam passive designs \citep{Sekiguchi2020,Rodriguez2018} tried to introduce a concrete benefit over the ADRs: the possibility of customizing a non-linear torque-angle curve depending on the user's needs. This advantage allows the adoption of alternative tailored designs to counteract for diverse pathological gaits and the allowance of some adaptation to different environments and settings. Although still lighter than active solutions, these devices generally are a bit heavier than other passive ones like the ADRs.

With the appearance of passive clutch-dependent devices \citep{VanHoorn2022,Yandell2019,Wang2020a,Liu2021,Hirai2006,Leclair2018,Collins2015,Pardoel2019} researchers made the next effort to gain more control over the timing of the delivered assistance within the gait cycle. This potential benefit has been further exploited by quasi-passive AAFOs with the aim of reaching higher torque levels and kinematic compatibility than with passive ones \citep{Kumar2020, Dezman2018, Wang2022, Zhang2015}. The shortcoming for clutch-based solutions is that normally, they provide assistance only on a single direction (plantarflexion or dorsiflexion).

Still, (quasi) passive devices in general have not been extensively tested to prove their good performance in adapting to different users and walking terrains.
On the contrary, active AAFOs have the major benefit of being able to provide large levels of assistance fully controllable in amplitude and timing, which enables them to adapt to different velocities or varying ground conditions \citep{Shorter2011, Yeung2017, Lerner2018, ISI:000510675700010, Meijneke2021, Chen2022}. However, their application is restricted to the performance of their controllers and actuators. For example, most of the devices that have a rigid transmission use position control rather than torque control. These devices are able to precisely control the final kinematics of the device, and therefore of the joint, but without considering the exerted torque on the subject, which can be dangerous if it is not properly controlled. They are mainly applied to patients who do not have control over their limbs and who are not able to oppose the movement \citep{Choi2018, ISI:000510675700010, Lee2021a, Weerasingha2018}. To alleviate this strict position tracking, some researchers used impedance control to allow some deviation from predefined trajectories \citep{Lee2021a} or by using low-transmission brushless DC motors, which have low back-drivability, allowing users to have safer interaction with the AAFO \citep{DeMiguel-Fernandez2022a, DeMiguel-Fernandez2022b, Nabipour2021}.

The stiffer the actuator, the more delicate and accurate the controller should be to track the user's motion or intention. Therefore, most of the active devices use compliant actuation that decreases the demands of the controllers due to the inherent adjustable behavior of the actuator. Starting from the geared actuator, the straightforward compliant system is formed by adding an elastic element, i.e. a spring, in series with the actuator \citep{Kim2011, VanDijk2017, Durandau2022}. These SEA actuators reduce the restriction imposed by the low back-drivability of the rigid transmission, being a compromise solution between maximal torque and weight that can fit well to assist impaired patients. However, depending on the device implementation, their size and bulkiness could not be yet suitable for daily use, so improving these is an open challenge.

An important issue about the assistance provided by active AAFOs is the strategy followed to deliver a timed assistance without disturbing the subject. In this sense, non-compliant devices with a rigid power transmission can only rely on control strategies that synchronize the subject's gait with the action of the AAFO. However, the nature of compliant mechanisms allows for a slight absorption of the hampering effect of imprecise controllers, which permits the transparent behaviour of the AAFO when the assistive torque is not given. 

Devices based on PAMs are characterized by the best compromise between exerted force and lightweight \citep{Galle2017, Kinnaird2009, Koller2015}. However, their slow dynamics only allow their use in subjects walking or moving at a slow velocity, i.e. impaired or elderly subjects. In addition, these devices are limited by their necessity of a compressed air supply to power them, which constrains their application to assist daily life activities.

Finally, AAFOs driven by cables have the potential of being the most lightweight active approach \citep{Slade2022, McCain2019, Bae2018}. By placing the actuator in a proximal location (mostly around the pelvis), the extra inertia added to the distal segment of the limb is reduced. This feature would enable them to be used in daily life scenarios, although they have not been extensively tested in such conditions. As a drawback, cable-driven AAFOs are only able to assist the motion in the traction direction of the cable, normally the plantarflexion direction of the ankle. This would prevent the assistance during the full gait cycle, but could be enough to assist certain gait functions such as push-off. Alternatively, by adding more than one cable and actuator, the device could assist both dorsi- and plantarflexion \citep{RewalkRobotics}.

\subsection{Primary goals with AAFOs when assisting gait}

By analyzing the experimental validations performed with the reviewed AAFOs (Figure~\ref{fig:experimentalValidation}), we can conclude that active devices have aroused more interest in the scientific community, dealing with a greater number of related articles. Moreover, none of the reviewed quasi-passive AAFOs have been finally validated with impaired subjects, even when they were designed to do so. This fact may imply that the readiness-level of quasi-passive solutions is lower than for purely passive or active devices and, therefore, further research effort should be done before this technology can be used in real-life scenarios by people with impairments. 

The push-off assistance is the preferred goal for all three AAFOs categories (i.e. passive, quasi-passive and active). In most cases, the target populations for this type of support are healthy (empowerment), or aging and people who suffer from muscle weaknesses (assistance). Most of these devices have only been validated with experiments on healthy subjects (except \citep{Sekiguchi2020} for passive, and \citep{ISI:000295363000003, Gomez-Vargas2021,Awad2017a,Tamburella2020,Takahashi2015,McCain2019,Harvey2021} for active). A reason for the limited evaluation might be that most of these devices require subjects with good balance capabilities as the devices do not normally account for that. For future use of these devices in populations with impairments it is key that they account for balance loss, assisting the user in that regard, or at least not being detrimental.

In the case of passive and quasi-passive devices, the capacity to support push-off is normally assessed by reductions in muscular activity. All the reviewed (quasi) passive studies reported reductions of plantarflexors muscular activity. However, dorsiflexors like the tibialis anterior, were normally not assessed or even showed an increase of activity \citep{Collins2015, Wang2022}. Only \citep{Kumar2020} reported a reduction in tibialis anterior and soleus at the same time, but their validation was only performed on a single healthy subject.These results suggest that extensive validations are still required for these devices.

Regarding active AAFOs, assisting push-off is the preferred objective for most cable-driven and PAM configurations. In this case, the validations focused on reducing the metabolic cost of walking in healthy subjects and improving kinematics and kinetics for paretic ankle propulsion. Based on these metrics, both approaches appeared to be valid alternatives that achieved their assistive purpose.

After push-off, the most common assistance is full-gait. For passive devices, ADRs and spring-cam are the approaches that allow for a double dorsi- and plantarflexion support \citep{Kerkum2014, Wren2015, kobayashi2017} but have their limitations. As mentioned before, the assistance provided is not timed within the gait cycle. This implies that instead of assisting the residual capabilities of the subject to perform a specific ankle motion, they provide stability while might restrict motion. Most of the ADR-AAFOs were tested with impaired subjects and some qualitative improvements in subjects' kinematics were reported. Nevertheless, no quantitative extensive discussion over their effects on the users is published (e.g. to evaluate if the added mass of the device to provoke those improvements is justified). Therefore, in spite of these promising results, there is not solid evidence of their benefits when assisting the complete gait cycle.

Active AAFOs offer more possibilities to assist full gait but for some mechanical designs, this comes at the cost of a double configuration. Only for the usually employed SEAs or rigid transmission the actuation provided in both dorsi- and plantarflexion is intrinsic to their design. However, PAM-based or cable-driven devices could also allow for that following a double configuration, with the actuator placed in parallel to both the agonist and antagonist muscle groups. In spite of this potential solution that would extend the capabilities of these groups of devices, the double configuration has only been explored by cable-driven AAFOs (e.g. \citep{Awad2017b, Gomez-Vargas2021, RewalkRobotics}).  

\subsection{Comprehensive analysis and need for extensive evaluations}
As mentioned earlier, due to the lack of consensus in the used experimental procedures between studies, it is very difficult to find common reported criteria for an objective comparison of the performance of different AAFOs.

When addressing the functional performance of their device, technological indicators that describe the physical capability of the AAFO are normally provided. In that regard, researchers tried to report results on, for example, joint kinematics, spatio-temporal parameters, interaction forces, adaptability to walking speed, and effects on muscular activity or energy consumption. However, not all these topics were addressed for all devices, and even more importantly, the topics contemplated were assessed following different methodologies or with different target populations. Therefore and in order to have a fair quantitative comparison, it is essential to make sets of bench-marking criteria that researchers can follow to evaluate their devices, which is the approach also proposed by Torricelli et al. \citep{Torricelli2020}.

Additionally, there is also a lack of evaluations of the devices in diverse conditions that are decisive for representing daily-life situations. Currently, devices are mostly evaluated during flat overground walking, treadmill, or just technically. Experiments on non-standardized settings like uneven terrains, when maneuvering around, or just start and stop gait for instance, have not been sufficiently discussed. Thus, although it is possible to hypothesize, for example, that AAFOs with body-weight dependent clutches are superior to spring-cam or angle-dependent clutches in providing assistance while walking on a slope, no evidence is available for proving this claim.

The effect of the AAFOs on users' balance or user's safety further than locomotion support are important aspects that are scarcely considered in the literature. There are some studies with active AAFOs that investigated the capacity of the device on improving balance capabilities of the users \citep{Gonzalez2022b, Bayon2022}, but they were not focused on assessing the mechanical configuration of the orthosis itself, but on evaluating a specific controller (except \citep{Harvey2021}).

Finally, the users' perceived experiences in terms of perceptual, emotional and cognitive aspects are also barely reported. A reason for that might be the small number of studies and evaluations with human beings at long term, which limits the inclusion of questionnaires or human-in-the-loop experiments with variations of the technical device that report experiences of final-users. These reports would be very useful for example to assess the perceived comfort, control (embodiment and agency) and required effort when using an specific AAFO.

Based on the above-mentioned points, the advantages and disadvantages reported in Section~\ref{sec:AdvDisadv}, and the data found on the literature (see Table S2 in the Supplementary Materials), we provide a subjective analysis of the clusters created in this article in terms of the readiness level we think they have in terms of functional performance and user experience (Tables \ref{table_FunctionalPerformance} and \ref{table_UserExperience} respectively). The features included in these tables are categorized using the framework of \citep{Torricelli2020}, and can also be used as a potential set of criteria that researchers can utilize when developing new devices in order to ensure a future consistent evaluation.

\renewcommand{\arraystretch}{1}
\setlength{\tabcolsep}{0.7em}
\begin{table}[h]
    \centering \small
    \caption{Features for functional performance and subjective evaluation of readiness level for each cluster. Grades are presented following a 5-points scale with (++) as very good, (+) as good, (+/--) as acceptable, (--) as poor and (--) as very poor.}
     \begin{tabular}{p{0.09\textwidth}C{0.09\textwidth}C{0.09\textwidth}C{0.09\textwidth}C{0.09\textwidth}C{0.09\textwidth}C{0.09\textwidth}}
        \\
         \rule{0pt}{15pt}
         & \multicolumn{6}{c}{\cellcolor{Gray}Functional performance} \\ [1.3ex]

        & \multirow{2}{*}{\parbox{0.09\textwidth}{Kinematic compatibility}} & \multirow{2}{*}{\parbox{0.09\textwidth}{Kinetic (dynamic) adjustability}} & \multirow{2}{*}{\parbox{0.09\textwidth}{Balance capabilities}} & \multirow{2}{*}{\parbox{0.09\textwidth}{Adaptation to non-standardized settings}} & \multirow{2}{*}{\parbox{0.09\textwidth}{Safety}} & \multirow{2}{*}{\parbox{0.09\textwidth}{Weight and other}} \\ [8ex] \cline{2-7}

        \rule{0pt}{10pt} ADR & +/- & -- --  & --   & -- --  & +   & +   \\[1.5ex]
        Spring-CAM & +/-- & +/-- & +/-- & --   & +   & +   \\[1.5ex]
        \rule{0pt}{15pt}Clutch & +/-- & +   & --   & +/-- & +   & ++  \\[1.5ex]
        Elect. Clutch & +   & +   & --   & +/-- & +/-- & +/-- \\[1.5ex]
        \rule{0pt}{15pt}Rigid transm. & +   & +   & +/-- & +/-- & --   & --   \\[1.5ex]
        \rule{0pt}{15pt}SEA & +   & ++  & +   & +   & +/-- & --   \\[1.5ex]
        Cable driven & +   & +   & --   & ++  & +/-- & +/-- \\
        \rule{0pt}{15pt}PAM & +   & +/-- & +/-- & +/-- & --   & +  \\[4ex]

     \end{tabular}
\label{table_FunctionalPerformance}
\end{table}

\begin{table}[ht]
    \centering \small
    \caption{Features for perceived users' experiences with the AAFOs and subjective evaluation of readiness level for each cluster. Grades are presented following a 5-points scale with (++) as very good, (+) as good, (+/--) as acceptable, (--) as poor and (--) as very poor.}
     \begin{tabular}{p{0.09\textwidth}C{0.15\textwidth}C{0.15\textwidth}C{0.15\textwidth}C{0.15\textwidth}}
        \\        
         \rule{0pt}{15pt} & \multicolumn{4}{c}{\cellcolor{Gray}User experience} \\ [1.3ex]

        & \multirow{2}{*}{\parbox{0.09\textwidth}{Comfortability}} & \multirow{2}{*}{\parbox{0.09\textwidth}{Sense of embodiment}} & \multirow{2}{*}{\parbox{0.09\textwidth}{Sense of agency}} & \multirow{2}{*}{\parbox{0.09\textwidth}{Cognitive effort}} \\ [4ex] \cline{2-5}

        \rule{0pt}{10pt} ADR & --   & -- -- & -- --  & +   \\[1.5ex]
        Spring-CAM & +/-- & --   & -- --  & +   \\[1.5ex]
        \rule{0pt}{15pt}Clutch & +   & +/-- & +/-- & +/-- \\[1.5ex]
        Elect. Clutch & +   & +/-- & +/-- & +/-- \\[1.5ex]
        \rule{0pt}{15pt}Rigid transm. & +/-- & --   & --   & --   \\[1.5ex]
        \rule{0pt}{15pt}SEA & +/-- & +   & +   & +   \\[1.5ex]
        Cable driven & +/-- & +   & +/-- & ++  \\[1.5ex]
        \rule{0pt}{15pt}PAM & --  & +/-- & +/-- & +/-- \\[4ex]
        
        
     \end{tabular}
\label{table_UserExperience}
\end{table}

\subsection{Future designs}
There have been quite some advances in passive AAFOs in the last years, but still these devices are not robustly prepared to address all kinds of situations encountered in everyday life. Started from less than 10 years ago, quasi-passive solutions represent a compromise between passive and active devices, resulting in an increasing trend that opted for using low electrical power to activate clutches. A trend for these devices (started in 2016) was to use other types of actuation rather than just springs and dampers for applying the assistance. The electroadhesive and dielectric elastomer actuators are examples of this. However, most of the passive and quasi-passive approaches have the limitation of providing assistance only towards dorsi- or plantarflexion, but not both. A groundbreaking step for (quasi) passive designs would be to have a device able to provide a configurable assistance on both directions with minor modifications.

Active AAFOs have the potential of being fully controllable to assist several gait functions with a single design. These devices are now evolving to lighter and more compliant solutions that should be transparent for the user while they are not providing assistance. Following the tendency of passive devices to assist activities of daily living, active AAFOs should also be tested in unrestricted environments to assess their performance when assisting gait under more challenging and rapidly varying conditions (variations of terrain and gait velocity).

\section*{Conflict of Interest Statement}

The authors declare that the research was conducted in the absence of any commercial or financial relationships that could be construed as a potential conflict of interest.

\section*{Author Contributions}

JLM and MN performed the literature review, collected the information, drafted and wrote the manuscript. EvA revised the manuscript and made substantial comments. CB designed the conceptualization and design of the study, contributed with the literature review, collected the information, drafted, wrote and reviewed the manuscript. All authors have read and approved the submitted version.

\section*{Funding}
This work has been carried out with the financial support from the Dutch Research Council NWO, under the grant Veni-TTW-2020 with ref. number 18079, and partly with the financial support from the KNAW Ter Meulen grant of the Academy Medical Sciences Fund, Royal Netherlands Academy of Arts \& Sciences, ref. number KNAWWF/1327/TMB202101.


\section*{Supplemental Data}
The supplementary material for this article can be found in SupplementaryMaterial file.


\bibliographystyle{Frontiers-Harvard} 
\bibliography{Main}

\newpage
\section*{Figure captions}


\clearpage

\end{document}


\onecolumn
\firstpage{1}

\title {{\helveticaitalic{Supplementary Material}}}

\maketitle

\section{Supplementary Tables}

\renewcommand{\arraystretch}{1.5}
\begin{table}[t]
\centering
\caption{Complement of Table 1 of the main manuscript. The publications are classified depending on the type of assistance and purpose (empowerment or rehabilitation) across passive, quasi-passive and active devices. Note than some devices are intended to support both impaired and healthy users.}

\resizebox{\textwidth}{!}{%
\label{table:purposeDistribution}

\begin{tabular}{M{2cm} M{2cm} M{1cm} M{2cm} M{2cm} M{2cm} M{2cm} M{2cm} M{2cm}}
\multicolumn{1}{l}{} &
  \multicolumn{1}{l}{} &
  \multicolumn{1}{l}{} &
  \multicolumn{6}{c}{Type of assistance} \\ \cline{4-9} 
 &
   &
   &
  Drop-foot &
  Foot-slap &
  Push-off &
  Full gait &
  Drop-foot + foot-slap &
  Drop-foot + push-off \\ \hline
\multicolumn{1}{c|}{\multirow{7}{*}{Passive}} & Impaired assist. & \textbf{11} & \citep{Rodriguez2018,Hirai2006} & \citep{Yamamoto2005,Yamamoto2022} & \citep{Sekiguchi2020, VanHoorn2022, Liu2021} & \citep{Kerkum2014,Kerkum2015, Meyns2020, Wren2015, UltraflexSystems2009, Kobayashi2019, Ottobock2018} &   &   \\
\multicolumn{1}{c|}{}                         & Healthy assist.  & \textbf{3}  &   &   & \citep{Wang2020a, Wiggin2011, Collins2015, Pardoel2019}
  &    &   &   \\
\multicolumn{1}{c|}{}                         & Both assist.     & \textbf{2}  &   &   & \citep{Yandell2019,Leclair2018}
  &    &   &   \\
\multicolumn{1}{c|}{}                         & Rehabilitation    & \textbf{}   &   &   &    &    &   &   \\ \cline{2-9} 
\multicolumn{1}{c|}{}                         & TOTAL    & \textbf{16} & 3 & 1 & 8  & 4  &   &   \\ \hline
\multicolumn{1}{c|}{\multirow{6}{*}{Quasi-passive}} &
  Impaired assist. &
  \textbf{2} &
  \citep{Allen2021} &
   &
   &
   &
  \citep{Oba2019a} &
   \\
\multicolumn{1}{c|}{}                         & Healthy assist.  & \textbf{3}  &   &   & \citep{Diller2016a,Kumar2020,Dezman2018}
  &    &   &   \\
\multicolumn{1}{c|}{}                         & Both assist.     & \textbf{1}  &   &   & \citep{Wang2022} &    &   &   \\
\multicolumn{1}{c|}{}                         & Rehabilitation    & \textbf{}   &   &   &    &    &   &   \\ \cline{2-9} 
\multicolumn{1}{c|}{}                         & TOTAL    & \textbf{6}  & 1 &   & 4  &    & 1 &   \\ \hline

\multicolumn{1}{c|}{\multirow{12}{*}{Active}}  & Impaired assist. & \textbf{19} & \citep{ISI:000220173200003,Kim2020} &   & \citep{Oymagil2007, ISI:000295363000003, Takahashi2015,McCain2019,Chen2022, Wang2022h} & \citep{Roy2009, Roy2013, ISI:000377575200001,Choi2018,Yeung2021, Yeung2018, Yeung2017,Lerner2018, Lerner2019a, Gasparri2019, Orekhov2020, Orekhov2021, Fang2021, Conner2020, Conner2021a, Fang2022a, Fang2022b,DeMiguel-Fernandez2022a, DeMiguel-Fernandez2022b,Chen2022d} & \citep{Kim2011} & \citep{Xia2020,Gomez-Vargas2020a, Gomez-Vargas2021,Bae2015, Awad2017a, Awad2017b, Bae2018, Siviy2020,Lee2021a} \\
\multicolumn{1}{c|}{}                         & Healthy assist.  & \textbf{10} &   &   & \citep{Mooney2014, Mooney2014a,Mooney2016,Peng2022b, Acosta-Sojo2022,Gonzalez2022b,Shepherd2022,Jackson2015, Witte2015, Steele2017,Witte2015, Zhang2017c, Jackson2019, Slade2022,Bougrinat2019,VanDijk2017, Dzeladini2016, Tamburella2020,Meijneke2014a,Gordon2007, Kinnaird2009, Cain2007, Kao2010b,Sawicki2008, Koller2015, Koller2017,Slade2022}
 &    &   &   \\
\multicolumn{1}{c|}{}                         & Both assist.    & \textbf{1}  &   &   &    & \citep{Bayon2022, Durandau2022, Meijneke2021} &   &   \\
\multicolumn{1}{c|}{}                         & Rehabilitation    & \textbf{2}  &   &   &    & \citep{Shorter2011, Li2011,Weerasingha2018} &   &   \\ \cline{2-9} 
\multicolumn{1}{c|}{}                         & TOTAL    & \textbf{32} & 2 &   & 15 & 10 & 1 & 4 \\ \hline
\end{tabular}%
}
\end{table}

\setlength{\tabcolsep}{0.4em} 
\begin{landscape}
\setlength\LTcapwidth{\linewidth} 
\begin{longtable}{p{1cm} p{1cm} p{1cm} p{1cm} p{1cm} p{2cm} p{2cm} p{2cm} p{4cm} p{3cm} p{3cm}}
\caption{Summary of reviewed passive AAFOs and their effects when being tested in subjects.}
\label{TablePassiveApend}\\
\hline
  \centering Ref. &
  \centering Assisted DOFs &
  \centering Cluster &
  \centering Weight (kg) &
  \centering Max. ROM (deg) &
  \centering Max. Assist. Spec. &
  \centering Assist. Type &
  \centering Subjects &
  \centering Kinematics &
  \centering EMG &
  \centering Energy consumption\cr \hline
\endhead
\citep{Yandell2019} &
  1/3 &
  BWC &
  0.459 &
  Full ROM &
  0.1-0.2 Nm/kg &
  Push-off &
  2 healthy &
  The exoskeleton adapts to changes in walking speeds &
  Reductions of 5-17\%   in average soleus activity comparing to no-exo condition &
  - \\
\citep{Kerkum2014,Kerkum2015,   Meyns2020} &
  1/1 &
  ADR &
  - &
  $\pm$15 &
  0.4 Nm/kg &
  Full gait &
  15 children with CP&
  Observed reduction on   knee flexion during stance similar to the one achieved by rigid AFOs. Worse   trunk performance and gait instability when wearing the ADR-AAFO &
  - &
  No   reduction \\
\citep{Wren2015,   UltraflexSystems2009} &
  1/1 &
  ADR &
  - &
  $\pm$40 &
  Adult $\pm$68Nm; Child $\pm$15.8 Nm &
  Full gait &
  10 children with CP&
  ADR produced better   knee extension and push-off power than PLS-AFO. PLS produced more normal   ankle motion, better walking and greater parent satisfaction &
  - &
  - \\
\citep{Kobayashi2019} &
  1/1 &
  ADR &
  - &
  $\pm$18 &
  $\pm$30 Nm &
  Full gait &
  10 stroke survivors &
  Ankle and knee   kinematics were affected by changes on the AAFO parameters (spring resistance,   zero alignment…) &
  - &
  - \\
\citep{Rodriguez2018} &
  1/1 &
  Spring-CAM &
  1.4 &
  +25 / -20 &
  - &
  Drop-foot &
  1 healthy &
  No changes &
  Reduced tibialis   anterior activity during swing phase (43.9\% of compensation) &
  - \\
\citep{Wang2020a} &
  1/1 &
  BWC &
  0.7655 &
  Full ROM &
  - &
  Push-off &
  5 healthy &
  The exo introduced   inertia that affected the foot positioning and clearance. Shorter stride   length. Drop foot occurred while wearing the device &
  Soleus activity reduced by 72\% &
  - \\
\citep{Wiggin2011,   Collins2015} &
  1/1 &
  AAC &
  0.408 - 0.503 &
  Full ROM &
  0.37 Nm/kg &
  Push-off &
  9 healthy &
  The exo did not interfere with normal kinematics &
  Soleus activity decreased with increasing spring stiffness. Gastroc. activity decreased during early and mid-stance, but increased during late stance. Tibialis anterior   activity increased during early and mid-stance, and ws unchanged during late stance &
  Reduction   of the metabolic cost of walking by 7.2$\pm$2.6\% for healthy human users \\
\citep{Ottobock2018} &
  1/1 &
  ADR &
  - &
  ?/-20 &
  - &
  Full gait &
  - &
  - &
  - &
  - \\
\citep{Sekiguchi2020} &
  1/1 &
  Spring-CAM &
  0.435 &
  Full ROM &
  - &
  Push-off &
  11 stroke survivors   with hemiparesis &
  The device (AAFO-PCAM) facilitated the realization of the ankle plantarflexion power in the   late-stance phase (compared to just AFO-P) because of dorsiflexion   resistance, increasing the knee flexion angle during the swing phase &
  - &
  - \\
\citep{Leclair2018} &
  1/1 &
  AAC &
  1.45 &
  +45 / -20 &
  115 Nm (technical tests) &
  Push-off &
  - &
  - &
  - &
  - \\
\citep{VanHoorn2022} &
  - &
  BWC &
  - &
  - &
  0.3 Nm/kg &
  Push-off &
  - &
  - &
  - &
  - \\
\citep{Hirai2006} &
  1/1 &
  BWC &
  - &
  0 / Full dorsiflexion &
  4.07 Nm &
  Drop-foot &
  1 healthy &
  Prevented drop foot when the subject tried to simulate it. The steps were longer with the AAFO than   without AAFO when simulating drop foot &
  - &
  - \\
\citep{Liu2021} &
  1/1 &
  BWC &
  - &
  Full ROM &
  - &
  Push-off &
  - &
  - &
  - &
  - \\
\citep{Pardoel2019} &
  1/1 &
  AAC &
  1.51 &
  Full ROM with   compliant PAM &
  - &
  Push-off &
  5 healthy &
  ROM of both ankles was   affected by wearing a unilateral device. The device minimally affected the   walking speed as well as the angular ROM of the knee and hip joints &
  Activation of the calf   muscles (Gas Med and Soleus) was reduced on the leg with the device &
  - \\
\citep{Yamamoto2005,Yamamoto2022} &
  1/1 &
  Other &
  0.1 (only mechanism) &
  +18/-30 &
  14 Nm &
  Foot-slap &
  17 stroke   survivors &
  More ankle   dorsiflexion in single stance and more peak power absorption in stance for   the AAFO-OD compared to a rigid AFO group. No difference found in peak   plantarfexion moment, ankle power generation, spatial or temporal parameters,   ground reaction force, or shank-to-vertical angle between the two groups &
  - &
  -\\[0.5ex]\hline
\multicolumn{11}{l}{\small BWC: body-weight dependent clutch; ADR: adjustable dynamic response; AAC: ankle-angle dependent clutch; Max. ROM: plantarflexion $>0$; dorsiflexion $<0$}

\end{longtable}
\end{landscape}
\clearpage

\setlength{\tabcolsep}{0.4em} 
\begin{landscape}
\setlength\LTcapwidth{\linewidth} 
\begin{longtable}{p{1cm} p{1cm} p{1cm} p{1cm} p{1cm} p{2cm} p{2cm} p{2cm} p{4cm} p{3cm} p{3cm}}
\caption{Summary of reviewed quasi-passive AAFOs and their effects when being tested in subjects.}
\label{TableQuasiPassiveApend}\\
\hline
  \centering Ref. &
  \centering Assisted DOFs &
  \centering Cluster &
  \centering Weight (kg) &
  \centering Max. ROM (deg) &
  \centering Max. Assist. Spec. &
  \centering Assist. Type &
  \centering Subjects &
  \centering Kinematics &
  \centering EMG &
  \centering Energy consumption\cr \hline
\endhead
\citep{Wang2022} &
  1/1 &
  EC &
  0.754 &
  ?/-12 &
  0.27 Nm/kg &
  Push-off &
  6 healthy &
  No changes &
  Level ground:   reductions in SOL (6.8$\pm$1.4\%) and GAS (6.6$\pm$1.0\%).   Increase in TA (4.4$\pm$0.8\%) & 
  Level   ground: reductions of 6.4$\pm$1.3\%;      - Ramp up: reductions of 5.2$\pm$1.2\% at 10\% grade and of 4.1$\pm$1.1\%   at 15\% grade; Ramp down: increase of 2.6$\pm$1.0\% at 10\% grade and by 2.4 ± 0.5\% at   15\% grade\\
\citep{Zhang2015} &
  1/3 &
  EC &
  - &
  +29/-19 &
  120 Nm &
  Push-off &
  - &
  - &
  - &
  - \\
\citep{Allen2021} &
  1/1 &
  Other &
  1.3 &
  - &
  7.8 N &
  Drop-foot &
  - &
  - &
  - &
  - \\
\citep{Diller2016a} &
  1/1 &
  Other &
  - &
  $\pm$30 &
  7.37 Nm &
  Push-off &
  1 healthy &
  - &
  - &
  - \\
\citep{Oba2019a} &
  1/1 &
  Other &
  0.819 &
  +20/-? &
  1.88 Nm &
  Drop-foot and   foot-slap &
  7 healthy &
   &
  - &
  - \\
\citep{Kumar2020} &
  1/1 &
  Other &
  1.75 &
  - &
  362 Nm/rad &
  Push-off &
  1 healthy &
  - &
  Reductions of 26.48\%   for TA and 7.42\% for SOL &
  - \\
\citep{Dezman2018} &
  1/1 &
  EC &
  1.09 &
  $\pm$10 &
  - &
  Push-off &
  7 healthy &
  - &
  - &
  -\\[0.5ex]\hline
\multicolumn{11}{l}{\small EC: electrically-controlled clutch; Max. ROM: plantarflexion $>0$; dorsiflexion $<0$}
\end{longtable}
\end{landscape}
\clearpage


\setlength{\tabcolsep}{0.4em} 
\begin{landscape}
\setlength\LTcapwidth{\linewidth} 
\begin{longtable}{p{1cm} p{1cm} p{1cm} p{1cm} p{1cm} p{1cm} p{1cm} p{3cm} p{2cm} p{4cm} p{3cm} p{2cm}}
\caption{Summary of reviewed active AAFOs and their effects when being tested in subjects.}
\label{TableActiveApend}\\
\hline
  \centering Ref. &
  \centering Assisted DOFs &
  \centering Cluster &
  \centering Weight (kg) &
  \centering Max. ROM (deg) &
  \centering Max. Assist. Spec. &
  \centering Assist. Type &
  \centering Control &
  \centering Subjects &
  \centering Kinematics &
  \centering EMG &
  \centering Energy consumption\cr \hline
\endhead
\citep{ISI:000220173200003} &
  1/1 &
  SEA &
  2.6 &
  - &
  - &
  Drop-foot &
  Impedance controller. Impedance level depends on gait state (swing/stance) &
  2 patients with drop-foot &
  Improvement of swing kinematics   and reduction of foot-slap &
  - &
  - \\
\citep{Kim2011} &
  1/1 &
  SEA &
  2.8 &
  +21.5/-12 &
  - &
  Drop-foot and foot-slap &
  The SEA is shortened or lengthened to induce dorsiflexion or plantarflexion. Torque direction according to the gait state &
  3 hemiparetic patients &
  Improvement of spatio-temporal parameters by  preventing   drop-foot &
  - &
  - \\
\citep{Roy2009,   Roy2013, ISI:000377575200001} &
  2/3 &
  SEA &
  3.6 &
  +45/-25 &
  23Nm &
  Full gait &
  Impedance controller - Variable   impedance and assistive torque profiles depending on biomechanical models and   current gait state &
  $>20$ patients &
  Reduced drop-foot &
  - &
  - \\
\citep{Choi2018} &
  1/1 &
  RT &
  0.869 &
  - &
  20Nm &
  Full gait &
  Position control &
  - &
  - &
  - &
  - \\
\citep{Oymagil2007,   ISI:000295363000003} &
  1/1 &
  SEA &
  0.95 (only actuator) &
  - &
  110W &
  Push-off &
  Position control for spring   compression: Assistive profile triggered by heel-strike and scaled to   previous step duration \citep{Oymagil2007}; Storage/release of energy depending   on the gait state \citep{ISI:000295363000003} &
  3 stroke survivors &
  \citep{ISI:000295363000003} gait   features improvements &
  - &
  - \\
\citep{Kim2020} &
  1/1 &
  RT &
  2.6 &
  +45/-20 &
  9.8Nm &
  Drop-foot &
  Torque control. Finite state machine to detect swing/stance. Free motion during stance, assistive torque during swing &
  5 hemiparetic patients &
  Ankle dorsiflexion peak improved &
  - &
  - \\
\citep{Mooney2014,   Mooney2014a,Mooney2016} &
  1/1 &
  Cable driven &
  1.06 &
  - &
  2W/Kg &
  Push-off &
  Torque control. Assistive profile triggered by heel-strike &
  $>20$ healthy &
  The provided mechanical power replaces power at biological joints &
  - &
  Metabolic cost reduced \\
\citep{Peng2022b, Acosta-Sojo2022,Gonzalez2022b,Shepherd2022} &
  1/1 &
  Cable driven &
  1.5 &
  - &
  30Nm &
  Push-off &
  Torque control. Assistive   profile triggered by heel-strike (deep learning model for gait state   estimation in \citep{Shepherd2022}) &
  $>20$ healthy &
  \citep{Gonzalez2022b} participants   changed the manner in which they maintained stability while on a beam   (changes in the normalized stride length and duration), although overall   balance was not affected; \citep{Peng2022b} Individuals exhibited different sensitivity   towards actuation timing &
  \citep{Acosta-Sojo2022} 60\%   subjects reduced GM and 80\% increased TA activity; muscular response highly   varied between subjects &
  \citep{Shepherd2022} 5.2\%   decrease in metabolic expenditure \\
\citep{Shorter2011,   Li2011} &
  1/1 &
  RT &
  1.9 &
  - &
  9Nm &
  Full gait &
 Non-regulated torque control. \citep{Shorter2011} Footswitch signals trigger the torque application;\citep{Li2011} Continuous phase estimation based on cross-correlation with a   learned model &
  - &
  - &
  - &
  - \\
\citep{Yeung2021,   Yeung2018, Yeung2017} &
  1/1 &
  RT &
  1 &
  +30/-20 &
  16.7Nm &
  Full gait &
  Torque control. State machine to detect gait states and execute different motor profiles &
  3-14    stroke survivors &
  Drop-foot and foot-slap reduced, improved gait pattern and speed &
  - &
  - \\
\citep{Xia2020} &
  1/1 &
  Cable driven &
  1 &
  - &
  7.3Nm &
  Drop-foot and push-off &
  Torque control. Torque/force reference triggered by gait events &
  5 healthy &
  Confirmed feasibility for   foot-drop prevention and propulsion assistance &
  Confirmed feasibility for   drop-foot prevention and propulsion assistance &
  - \\
\citep{Lerner2018,   Lerner2019a, Gasparri2019, Orekhov2020, Orekhov2021, Fang2021, Conner2020,   Conner2021a, Fang2022a, Fang2022b} &
  1/1 &
  Cable driven &
  1.85-2.2 &
  - &
  20Nm &
  Full gait &
  Torque control. \citep{Lerner2018} Assistive torque profile triggered by gait events detection; \citep{Lerner2019a, Orekhov2021,Fang2022a, Fang2022b} assistance proportional to ankle moment during stance;   \citep{Gasparri2019, Orekhov2020, Fang2021} assistance proportional to ankle moment; \citep{Conner2020} resistance proportional to ankle moment &
  $>20$ patients with CP &
  Improved ankle plantarflexion and   knee extension, producing a more efficient gait pattern, improved 6-min   walking test performance, more symmetrical gait pattern with bilateral   assistance &
  Main reduction on soleus   activity &
  Metabolic cost reduced \\
\citep{Gomez-Vargas2020a,   Gomez-Vargas2021} &
  1/1 &
  Cable driven &
  2.8 &
  - &
  4.5Nm &
  Drop-foot and push-off &
  Torque control. Gait phases   detected by inertial sensor placed on the foot that triggers the assistance &
  10 stroke survivors &
  Biomechanical improvements in   70\% patients &
  - &
  - \\
\citep{Jackson2015,   Witte2015, Steele2017} &
  1/1 &
  Cable driven &
  0.88 &
  +30/-20 &
  119Nm &
  Push-off &
  Torque or work control. Torque profile triggered by heel strike and scaled to last step duration &
  $>20$ healthy &
  - &
  Reduced muscular activity and   effect on muscular coordination &
  Metabolic cost reduced with work   assistance but increased with torque assistance \\
\citep{Witte2015,   Zhang2017c, Jackson2019, Slade2022} &
  1/1 &
  Cable driven &
  0.88 &
  +30/-20 &
  121Nm &
  Push-off &
  Torque or work control. Torque profile triggered by heel strike and scaled to last step duration &
  $>20$ healthy &
  - &
  Reduced  soleus activity &
  Human-in-the-loop (EMG or torque   optimized) reduced metabolic cost \\
\citep{Galle2013,   Galle2014, Galle2015, Galle2017,Malcolm2018} &
  1/1 &
  PAM &
  0.89 &
  ?/-15 &
  - &
  Push-off &
  Work control. Torque profile triggered by heel strike and scaled to last step duration &
  $>20$ healthy &
  Walking performance increased (subjects were able to carry more weight) &
  Reduced leg muscular activity &
  Metabolic cost reduced \\
\citep{ISI:000510675700010} &
  1/1 &
  RT &
  - &
  - &
  - &
  Full gait &
  Position tracking by adaptive   proxy-based sliding mode control. Trajectory generation depending on gait   phase estimation based on time between key-events &
  Technically validated &
  - &
  - &
  - \\
\citep{Bae2015,   Awad2017a, Awad2017b, Bae2018, Siviy2020} &
  1/3 &
  Cable driven &
  3.8 &
  Full ROM &
  300N &
  Drop-foot and push-off &
\citep{Bae2015, Awad2017a,   Awad2017b} cable position control; 
\citep{Bae2018} iterative Force-based cable trajectory tracking;    \citep{Siviy2020} offline human-in-the-loop optimized force profiles tracked   by an admittance control &
  $>20$ stroke survivors &
  Improved gait symmetry, reduced hip hiking and circumduction &
  - &
  Metabolic cost reduced \\
\citep{Bougrinat2019} &
  1/1 &
  Cable driven &
  0.348 &
  Full ROM &
  30 Nm &
  Push-off &
  Open-loop torque control. Gait phase estimation from the duration of previous steps. Plantar-flexion torque  profile triggered at certain phase &
  1 healthy subject &
  - &
  Gastrocnemius activity decreased &
  - \\
\citep{VanDijk2017,   Dzeladini2016, Tamburella2020,Meijneke2014a} &
  1/1 &
  SEA &
  1.5 &
  - &
  100N &
  Push-off &
Torque control tracking a spring   simulated torque profile; \citep{VanDijk2017} Gait phase estimated by an AO based on the foot pressure; \citep{Dzeladini2016, Tamburella2020} Central Pattern Generator based on   floor contact events &
  \citep{VanDijk2017} 4 healthy   subjects; \citep{Tamburella2020} 4 incomplete spinal cord injury patients   during 10-session training &
  Improved gait speed and endurance   during assisted walking &
  Reduced muscle activity & \citep{VanDijk2017} No difference   in metabolic cost; \citep{Dzeladini2016} reduced metabolic cost of walking\\
\citep{Gordon2007,   Kinnaird2009, Cain2007, Kao2010b} &
  1/1 &
  PAM &
  1.2 &
  - &
  60Nm &
  Push-off &
  Plantarflexor torque control. \citep{Gordon2007} proportional myoelectric control from soleus EMG; \citep{ Kinnaird2009} Proportional myoelectric control from Medial   Gastrocnemius EMG;\citep{Cain2007} proportional soleus myoelectric control vs footswitch on/off control;\citep{Kao2010b} Proportional myoelectric    control from soleus EMG with fixed gain &
  $>20$ healthy &
  Kinematics closer to normal with myoelectric control. Differences in kinematic patterns but similar ankle   torque pattern with and without exoskeleton. &
  Main reduction on soleus   activity &
  - \\
\citep{Sawicki2008,   Koller2015, Koller2017} &
  1/1 &
  PAM &
  1.2 &
  - &
  60Nm &
  Push-off &
 Plantarflexor torque control. \citep{Sawicki2008} proportional myoelectric    control from soleus EMG with fixed gain; \citep{ Koller2015} proportional myoelectric    control from soleus EMG with adaptive gain; \citep{Koller2017} proportional myoelectric    control vs torque profile triggered by heel strikes &
  $>20$ healthy &
  Similar to normal ankle pattern;   reduced ankle power &
  Reduced soleus activity &
  Metabolic cost reduced \\
\citep{Takahashi2015} &
  1/1 &
  PAM &
  0.53 &
  - &
  - &
  Push-off &
  Plantarflexor torque control. Proportional myoelectric propulsion from soleus EMG &
  5 stroke survivors &
  Increased paretic plantarflexor torque &
  - &
  - \\
\citep{McCain2019} &
  1/1 &
  Cable driven &
  - &
  - &
  - &
  Push-off &
  Force control. Proportional   myoelectric propulsion from soleus EMG increased by gait velocity &
  6 stroke survivors &
  Increased paretic plantarflexor   torque &
  - &
  - \\
\citep{Bayon2022,   Durandau2022, Meijneke2021} &
  1/2 &
  SEA &
  3.8 &
  - &
  102Nm &
  Full gait &
 Torque controller. \citep{Meijneke2021} neuromuscular controller;      \citep{Bayon2022} balance recovering controller; \citep{Durandau2022} neuromechanical models to estimate biological ankle   joint torques in real-time from measured EMG and joint angles &
  3 SCI subjects; 16 healthy &
  SCI patients achieved   healthy-like kinematic patterns, balance improvements, reduced biological   torques &
  Reduced muscular activity   (soleus and medial gastrocnemius for recovering balance) &
  - \\
\citep{Chen2022,   Wang2022h} &
  1/1 &
  Cable driven &
  0.95-1.4 &
  - &
  50Nm &
  Push-off &
  \citep{Chen2022} Gait phase   estimation to generate torque profile; \citep{Wang2022h} Study about performance of nine different torque profiles &
  3-5 healthy subjects &
  Exoskeleton assistance changed   maximum ankle dorsiflexion and plantarflexion angle and reduced biological   ankle moment &
  Soleus activity decreased &
  Metabolic cost reduced \\
\citep{DeMiguel-Fernandez2022a,   DeMiguel-Fernandez2022b} &
  1/1 &
  RT &
  1.56 &
  - &
  30Nm &
  Full gait &
  Torque controller. Torque   profile triggered by heel strike and scaled to previous steps duration &
  5 stroke survivors &
  Drop-foot correction and reduced   hip compensatory movements &
  - &
  - \\
\citep{Slade2022} &
  1/1 &
  Cable driven &
  1.2 &
  +55/? &
  54Nm &
  Push-off &
  Speed-adaptive torque controller   optimized by AI &
  10 healthy subjects &
  Increased gait speed &
  - &
  Metabolic cost reduced \\
\citep{Choi2020} &
  2/2 &
  PAM &
  2.14 &
  - &
  - &
  Push-off &
  Force/Position control &
  7 healthy subjects &
  Increase balance of   inversion/eversion of the ankle &
  The EMG rate decreased by 17.1\% ± 11.3\% by PAM1 &
  - \\
\citep{Lee2021a} &
  1/2 &
  RT &
  2.1 &
  +50/-20 &
  12Nm &
  Drop-foot and push-off &
  Impedance control dependent from   gait state &
  1 healthy subject &
  Increased plantar and dorsiflexion are clearly shown &
  - &
  - \\
\citep{Weerasingha2018} &
  2/3 &
  RT &
  1.355 &
  +50/-20 &
  21Nm &
  Full Gait &
  Position control &
  1 healthy subjects &
   &
  Decrease of tibialis anterior and gastrocnemius peak EMG &
   \\
\citep{Chen2022d} &
  1/1 &
  SEA &
  3.05 &
  +40/-20 &
  Driving 22.6Nm; Braking 41.9 Nm &
  Full gait &
  Impedance control dependent from   gait state &
  8 healthy subjects & -
   &
  Reduction of tibialis anterior and soleus EMG & - \\[0.5ex]\hline
\multicolumn{11}{l}{\small RT: rigid transmission; SEA: series elastic actuator; PAM: pneumatic artificial muscle; Max. ROM: plantarflexion $>0$; dorsiflexion $<0$}

\end{longtable}
\end{landscape}
\clearpage





\bibliographystyle{unsrt} 
\bibliography{References}